\documentclass[aps,prl,twocolumn,showpacs,superscriptaddress]{revtex4-1}

\usepackage{graphicx}
\usepackage[utf8]{inputenc}
\usepackage{caption}
\usepackage{siunitx}
\usepackage{amsmath}
\usepackage{gensymb}
\usepackage{amssymb}
\usepackage{enumerate}
\usepackage{color}
\usepackage{hyperref}
\usepackage[american]{babel}
\usepackage{subcaption}
\usepackage{etoolbox}
\usepackage{floatrow}
\usepackage{nicefrac}

\patchcmd{\thebibliography}{\section*{\refname}}{}{}{}

\newcommand{\fig}[1] {Fig.~\ref{fig:#1}}

\newcommand{\pos}{$\mathrm{e^+}$}
\newcommand{\ele}{$\mathrm{e^-}$}

\newcommand{\orb}[3]{$#1^{#2}\text{#3}$}
\newcommand{\ttP}[0]{\orb{3}{3}{P}}

\newcommand{\ooS}[0]{\orb{1}{1}{S}}

\newcommand{\otS}[0]{\orb{1}{3}{S}}

\newcommand{\tS}[0]{\orb{2}{3}{S}}
\newcommand{\tP}[0]{\orb{2}{3}{P}}

\newcommand{\new}[1]{#1}

\DeclareSIUnit\inch{in.}
\DeclareSIUnit\division{div}

\begin{document}

\preprint{APS/123-QED}

\title{Efficient $ \mathrm{2^3S} $ positronium production by stimulated decay from the $ \mathrm{3^3P} $ level}

\newcommand{\corresponding}[1]{\altaffiliation{Corresponding author, #1}}

\newcommand{\affpolimi}[0]{\affiliation{LNESS, Department of Physics, Politecnico di Milano, via Anzani 42, 22100~Como, Italy}}
\newcommand{\affinfnmi}[0]{\affiliation{INFN, Sezione di Milano, via Celoria 16, 20133~Milano, Italy}}
\newcommand{\affvienna}[0]{\affiliation{Stefan Meyer Institute for Subatomic Physics, Austrian Academy of Sciences, Boltzmanngasse 3, 1090~Vienna, Austria}}
\newcommand{\affinsubria}[0]{\affiliation{Department of Science and High Technology, University of Insubria, Via Valleggio 11, 22100~Como, Italy}}
\newcommand{\affjinr}[0]{\affiliation{Joint Institute for Nuclear Research, Dubna~141980, Russia}}
\newcommand{\affbs}[0]{\affiliation{Department of Mechanical and Industrial Engineering, University of Brescia, via Branze 38, 25123~Brescia, Italy}}
\newcommand{\affinfnpv}[0]{\affiliation{INFN Pavia, via Bassi 6, 27100~Pavia, Italy}}
\newcommand{\afftn}[0]{\affiliation{Department of Physics, University of Trento, via Sommarive 14, 38123~Povo, Trento, Italy}}
\newcommand{\affinfntn}[0]{\affiliation{TIFPA/INFN Trento, via Sommarive 14, 38123~Povo, Trento, Italy}}
\newcommand{\affge}[0]{\affiliation{Department of Physics, University of Genova, via Dodecaneso 33, 16146~Genova, Italy}}
\newcommand{\affinfnge}[0]{\affiliation{INFN Genova, via Dodecaneso 33, 16146~Genova, Italy}}
\newcommand{\affmi}[0]{\affiliation{Department of Physics ``Aldo Pontremoli'', Universit\`{a} degli Studi di Milano, via Celoria 16, 20133~Milano, Italy}}
\newcommand{\affmpi}[0]{\affiliation{Max Planck Institute for Nuclear Physics, Saupfercheckweg 1, 69117~Heidelberg, Germany}}
\newcommand{\afflac}[0]{\affiliation{Laboratoire Aim\'e Cotton, Universit\'e Paris-Sud, ENS Paris Saclay, CNRS, Universit\'e Paris-Saclay, 91405~Orsay Cedex, France}}
\newcommand{\affpolimiII}[0]{\affiliation{Department of Aerospace Science and Technology, Politecnico di Milano, via La Masa 34, 20156~Milano, Italy}}
\newcommand{\affheidelberg}[0]{\affiliation{Kirchhoff-Institute for Physics, Heidelberg University, Im Neuenheimer Feld 227, 69120~Heidelberg, Germany}}
\newcommand{\affcern}[0]{\affiliation{Physics Department, CERN, 1211~Geneva~23, Switzerland}}
\newcommand{\affoslo}[0]{\affiliation{Department of Physics, University of Oslo, Sem Saelandsvei 24, 0371~Oslo, Norway}}
\newcommand{\afflyon}[0]{\affiliation{Institute of Nuclear Physics, CNRS/IN2p3, University of Lyon 1, 69622~Villeurbanne, France}}
\newcommand{\affmoscow}[0]{\affiliation{Institute for Nuclear Research of the Russian Academy of Science, Moscow~117312, Russia}}
\newcommand{\affinfnpd}[0]{\affiliation{INFN Padova, via Marzolo 8, 35131~Padova, Italy}}
\newcommand{\affprague}[0]{\affiliation{Czech Technical University, Prague, Brehov\'a 7, 11519~Prague~1, Czech Republic}}
\newcommand{\affbo}[0]{\affiliation{University of Bologna, Viale Berti Pichat 6/2, 40126~Bologna, Italy}}
\newcommand{\affpv}[0]{\affiliation{Department of Physics, University of Pavia, via Bassi 6, 27100~Pavia, Italy}}
\newcommand{\affnorway}[0]{\affiliation{The Research Council of Norway, P.O. Box 564, 1327~Lysaker, Norway}}
\newcommand{\affheidelbergII}[0]{\affiliation{Department of Physics, Heidelberg University, Im Neuenheimer Feld 226, 69120~Heidelberg, Germany}}
\newcommand{\affbsII}[0]{\affiliation{Department of Civil, Environmental, Architectural Engineering and Mathematics, University of Brescia, via Branze 43, 25123~Brescia, Italy}}



\author{M.~Antonello}
\affinfnmi
\affinsubria

\author{A.~Belov}
\affmoscow

\author{G.~Bonomi}
\affbs
\affinfnpv

\author{R.~S.~Brusa}
\afftn
\affinfntn

\author{M.~Caccia}
\affinfnmi
\affinsubria

\author{A.~Camper}
\affcern

\author{R.~Caravita}
\corresponding{ruggero.caravita@cern.ch}
\affcern

\author{F.~Castelli}
\affinfnmi
\affmi

\author{G.~Cerchiari}
\affmpi

%

\author{D.~Comparat}
\afflac

\author{G.~Consolati}
\affpolimiII
\affinfnmi

\author{A.~Demetrio}
\affheidelberg

\author{L.~Di~Noto}
\affge
\affinfnge

\author{M.~Doser}
\affcern


\author{M.~Fan\`{i}}
\affge
\affinfnge
\affcern




\author{S.~Gerber}
\affcern


\author{A.~Gligorova}
\affvienna

\author{F.~Guatieri}
\afftn
\affinfntn

\author{P.~Hackstock}
\affvienna

\author{S.~Haider}
\affcern

\author{A.~Hinterberger}
\affcern


\author{A.~Kellerbauer}
\affmpi

\author{O.~Khalidova}
\affcern

\author{D.~Krasnick\'y}
\affinfnge

\author{V.~Lagomarsino}
\affge
\affinfnge


\author{P.~Lebrun}
\afflyon

\author{C.~Malbrunot}
\affcern
\affvienna

\author{S.~Mariazzi}
\afftn
\affinfntn


\author{V.~Matveev}
\affmoscow
\affjinr


\author{S.~R.~M\"{u}ller}
\affheidelberg

\author{G.~Nebbia}
\affinfnpd

\author{P.~Nedelec}
\afflyon

\author{M.~Oberthaler}
\affheidelberg

\author{E.~Oswald}
\affcern

\author{D.~Pagano}
\affbs
\affinfnpv

\author{L.~Penasa}
\afftn
\affinfntn

\author{V.~Petracek}
\affprague

\author{F.~Prelz}
\affinfnmi


\author{B.~Rienaecker}
\affcern

\author{J.~Robert}
\afflac

\author{O.~M.~R{\o}hne}
\affoslo

\author{A.~Rotondi}
\affinfnpv
\affpv

\author{H.~Sandaker}
\affoslo

\author{R.~Santoro}
\affinfnmi
\affinsubria



\author{G.~Testera}
\affinfnge

\author{I.~C.~Tietje}
\affcern


\author{E.~Widmann}
\affvienna

\author{T.~Wolz}
\affcern

\author{P.~Yzombard}
\affmpi

\author{C.~Zimmer}
\affcern
\affmpi
\affheidelbergII


\author{N.~Zurlo}
\affinfnpv
\affbsII

\collaboration{The AEgIS collaboration}
\noaffiliation{}


\date{\today}

\pacs{32.80.Rm, 36.10.Dr, 78.70.Bj}

\begin{abstract}
We investigate experimentally the possibility of enhancing the production of \tS{} positronium atoms by driving the \otS{}--\ttP{} and \ttP{}--\tS{} transitions, overcoming the natural branching ratio limitation of spontaneous decay from \ttP{} to \tS{}. The decay of \ttP{} positronium atoms towards the \tS{} level has been efficiently stimulated by a \SI{1312.2}{\nano\meter} broadband IR laser pulse. The dependence of the stimulating transition efficiency on the intensity of the IR pulse has been measured to find the optimal enhancement conditions. A maximum relative increase of $ \times (3.1 \pm 1.0) $ in the \tS{} production efficiency, with respect to the case where only spontaneous decay is present, was obtained.
\end{abstract}

\maketitle{}


\new{Positronium (Ps) is the neutral matter-antimatter bound state of an electron (\ele{}) and a positron (\pos{}). Ps has two distinct ground states: the singlet \ooS{} (para-Ps), annihilating into two $\gamma$-rays with a lifetime of \SI{0.125}{\nano\second}, and the triplet \otS{} (ortho-Ps), annihilating into three $\gamma$-rays with a lifetime of \SI{142}{\nano\second} \cite{rich_review:81}. Ps, being a purely leptonic two-body system, is well-known for offering an ideal testing ground for high-precision Quantum Electrodynamics (QED) calculations \cite{karshenboim_rev:05}. Among the many precision experiments, the most accurate were recently conducted using two-photon Doppler-free laser spectroscopy of the \otS{}--\tS{} transition \cite{cassidy_review:18}. The \tS{} level has an extended lifetime of \SI{1142}{\nano\second} in vacuum. This is due to its optical metastability: single-photon radiative decays to \otS{} are prohibited by the electric dipole selection rules and the reduced overlap between the positron and the electron wave-functions increases its annihilation lifetime by a factor of eight \cite{charlton:01}. On top of its high-precision spectroscopy applications, \tS{} Ps is one of the few notable candidate systems being considered for measuring the gravitational interaction between matter and antimatter \cite{oberthaler_ps:02}, together with Ps in long-lived Rydberg states \cite{mills_leventhal:02, cassidy_psgravity:14}, antihydrogen \cite{alpha_grav:13, aegis_natc:14, gbar:15} and muonium \cite{mage:18}. 
Moreover, the metastable \tS{} Ps has a very low electrical polarizability, thus being scarcely sensitive to stray electric fields \cite{cassidy_ryd:15}, and is a good candidate for atom interferometry, provided that a beam with sufficiently low divergence and high intensity is demonstrated \cite{oberthaler_ps:02}. 
Furthermore, an intense source of polarized \tS{} atoms has been recently shown to be of extreme usefulness to achieve Bose-Einstein condensation of Ps \cite{zhang_metabec:19}.}
\par
\tS{} Ps sources have been demonstrated via RF transition from laser-excited \tP{} Ps in a weak magnetic field \cite{mills_23s:75}, via two-photon Doppler-free \otS{}--\tS{} laser excitation \cite{chu_1s2s_prl:84, fee_1s2s_prl:93} and more recently via single-photon excitation of \otS{} to \tP{} in a rapidly switching electric field \cite{alonso_metastable:17} and via single-photon excitation of \otS{} to \ttP{} with radiative decay to \tS{} in an electric field \cite{aegis_meta:18} and in absence of electric field \cite{aegis_meta2:19}. This last method in particular showed that it is possible to build an almost-monochromatic \tS{} Ps source with a selected and tunable velocity distribution in the $ 10^4 \, \si{\meter\per\second} $ range, with an overall efficiency between $ 0.7-1.4 \% $ according to the selected velocity \cite{aegis_meta2:19}.
\par
In the present work, following this experimental line, we investigate the possibility of stimulating the \ttP{}--\tS{} transition to increase the overall \tS{} production efficiency. Indeed, laser-excited \ttP{} Ps can spontaneously decay radiatively to \tS{} via the dipole-allowed \ttP{}--\tS{} transition (rate $ A_{23} = 2\pi \times 1.1 \cdot 10^7 \,\, \si{\per\second} $) with $ \sim \mkern-4mu 10 \% $ measured branching efficiency \cite{aegis_meta:18, aegis_meta2:19}, limited by the competition with the more efficient spontaneous decay channel \ttP{}--\otS{} ($ A_{13} = 2\pi \times 8.4 \cdot 10^7 \,\, \si{\per\second} $). Increasing the \ttP{}--\tS{} transition rate, and thus the branching efficiency of the \tS{} decay, is possible through stimulated emission.
\par
\begin{figure}[htp]
	\centering
	\includegraphics[width=0.9 \textwidth]{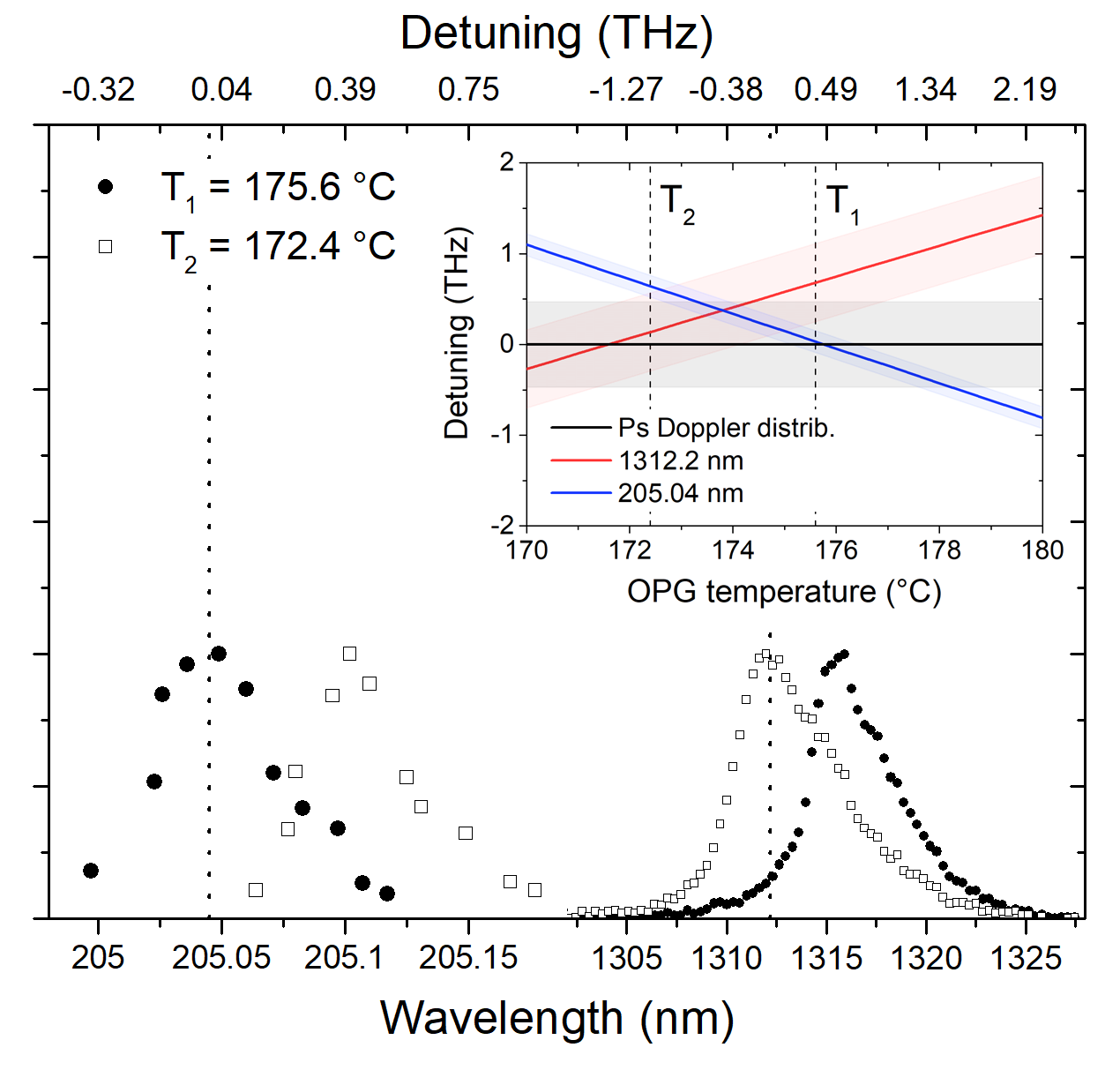}
	\caption{Laser spectra (arbitrary units) measured in the two OPG temperature set-points optimized for stimulating the \otS{}--\ttP{}--\tS{} transition (white squares, \SI{172.4}{\celsius}) and for the highest \otS{}--\ttP{} excitation efficiency (black circles, \SI{175.6}{\celsius}). The dotted lines mark the transition resonances. Inset: measured laser detuning from the theoretical resonance frequency as a function of the OPG crystal temperature for the \otS{}--\ttP{} \SI{205}{\nano\meter} laser (blue band) and the \ttP{}--\tS{} \SI{1312}{\nano\meter} laser (red band) respectively, compared to the reference Ps Doppler distribution as measured in \cite{aegis_neq3:16} (gray band).}
	\label{fig:detuning_laser}
\end{figure}

A straightforward way to stimulate the \ttP{}--\tS{} transition consists in introducing a synchronized broadband IR laser pulse at \SI{1312.2}{\nano\meter} \cite{oberthaler_ps:02} on top of the pulsed UV laser at $ \sim \SI{205}{\nano\meter} $ used for \otS{}--\ttP{} excitation \cite{aegis_neq3:16}. \new{In our experiment, such a laser pulse could be obtained from the same optical setup producing the UV beam, which is described in \cite{cialdi_nimb:11, master_caravita:13}. In this setup, a sequence of non-linear optical conversion processes are used to generate the \SI{205}{\nano\meter} wavelength, and in particular an optical parametric generation (OPG) crystal yields as a byproduct a broadband, amplified idler beam at \SI{1314}{\nano\meter}.} 
Both \SI{205}{\nano\meter} and \SI{1314}{\nano\meter} wavelengths could be tuned by varying the temperature set-point of the OPG crystal around its nominal working point of $ T_1 = \SI{175.6}{\celsius} $ (\fig{detuning_laser}). This temperature value, chosen for the UV wavelength to fall on the \otS{}--\ttP{} resonance ($ \lambda_3 = \SI{205.045}{\nano\meter} $), is ineffective for a stimulated emission experiment as the bandwidth of the IR does not arrive to cover the \SI{1312.2}{\nano\meter} wavelength with sufficient optical power to efficiently stimulate the \ttP{}--\tS{} transition. Hence, a different set-point $ T_2 = \SI{172.4}{\celsius} $ was selected to gain enough power at \SI{1312.2}{\nano\meter}, while mantaining an acceptable detuning $ \delta \nu_3 $ of the UV beam from the \otS{}--\ttP{} resonance frequency, to minimize the unavoidable reduction in excitation efficiency.
\new{The set-point setting accuracy was limited to \SI{0.5}{\celsius} by the TC200 temperature controller. At $ T_2 $, the induced \otS{}--\ttP{} detuning was $ {\delta \nu}_3 \approx \mkern-4mu \SI{410}{\giga\hertz} $, corresponding to \SI{205.103}{\nano\meter} and $ 0.9 \sigma $ of the Doppler distribution of our Ps source ($ \sigma_\nu \approx \mkern-4mu \SI{470}{\giga\hertz} $ \cite{aegis_neq3:16}). A reduction in the \otS{}--\ttP{} excitation efficiency of about $ \sim 50 \% $ was expected \cite{aegis_neq3:16}, due to the Doppler selection of a Ps atoms distribution with $ \lambda_3 {\delta \nu}_3 \approx 0.85 \cdot 10^5 \, \si{\meter\per\second} $ average velocity in the direction parallel to the laser beam.}

\begin{figure}[htp]
	\centering
	\includegraphics[width=\textwidth]{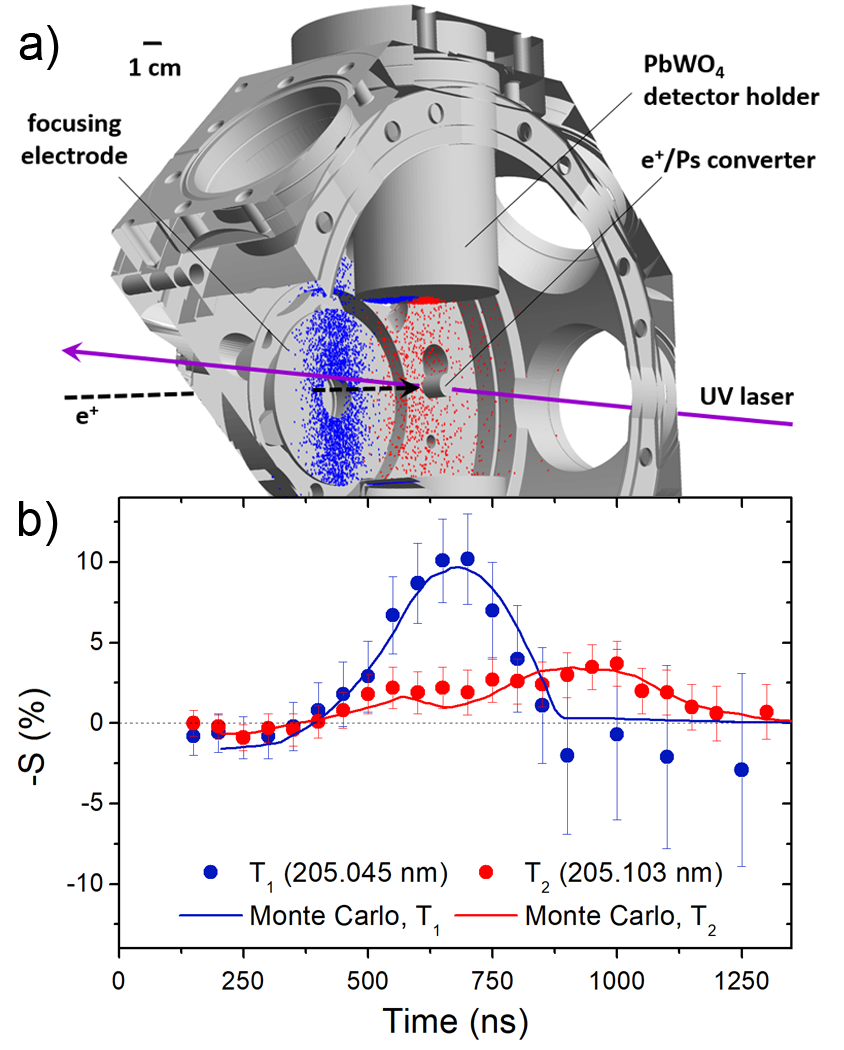}
	\caption{a) Distributions of Ps impact positions, shown superimposed to the 3D drawing of the chamber walls, for the $ T_1 = \SI{175.6}{\celsius} $ (blue circles) and $ T_2 = \SI{172.4}{\celsius} $ (red circles) setpoints, emphasizing the different Doppler selection in the two detuning conditions. b) Measurement of the annihilation time distributions of spontaneously-decaying \tS{} Ps atoms from the \ttP{} level without stimulated transition into \tS{} for the two temperature set-points. The graph shows the curve $ -S(t) = (A^\mathrm{UV}(t) - A^\mathrm{off}(t))/A^\mathrm{off}(t) $ (see text). Each time distribution has been fitted with the Monte Carlo model discussed in the text (solid lines).}
	\label{fig:setpoint_comparison}
\end{figure}

A direct way to observe an enhancement in the \tS{} signal due to the action of the stimulating laser is to compare the Ps annihilation time distribution with the UV laser only to that measured with both UV and IR pulses. The experimental methodology was the same used in previous works \cite{aegis_neq3:16, aegis_meta:18, aegis_meta2:19}. Bursts of ~$10^7$ \pos{}, \SI{7}{\nano\second} in time length \cite{aegis_nimb:15}, were guided by a \SI{25}{\milli\tesla} magnetic field, focused by an electric field of about \SI{300}{\volt\per\centi\meter} and implanted at \SI{3.3}{\kilo\electronvolt} in a circular spot of $ \sim $ \SI{3}{\milli\meter} full-width at half maximum (FWHM) into a \pos{}/Ps converter held at room temperature. The converter is constitued by a Si(111) p-type crystal with nanochannels produced via electrochemical etching and oxidized in air at \SI{100}{\celsius} for \SI{2}{\hour} \cite{mariazzi_prb:10}. Ps formed in the converter out-diffuses back into vacuum through the nanochannels with an efficiency up to ~35\% loosing a fraction of its formation energy by collisions with the channel walls \cite{mariazzi_prl:10}. A fraction of the emitted cloud was subsequently conveyed to \tS{} either by the \SI{205}{\nano\meter} UV beam alone (i.e. through spontaneous decay from \ttP{}, as in \cite{aegis_meta:18, aegis_meta2:19}) or by a combination of the UV and IR laser pulses (i.e. through stimulated decay from \ttP{}).
\new{Both beams were linearly polarized perpendicularly to the target, with a nearly Gaussian temporal profile with a FWHM of \SI{1.5}{\nano\second} (UV) and \SI{4.0}{\nano\second} (IR), and a nearly-Gaussian spectral profile with bandwidths $ \sigma^\mathrm{UV} \approx 2\pi \times \SI{120}{\giga\hertz} $ (UV) and $ \sigma^\mathrm{IR} \approx 2\pi \times \SI{440}{\giga\hertz} $ (\fig{detuning_laser}). \new{The energy of the two pulses was $ (53 \pm 5) \, \si{\micro\joule} $ for the UV and $ (405 \pm 10) \, \si{\micro\joule} $ for the IR (at the entrance viewport of the experimental chamber).} The UV spot was nearly Gaussian with $ (7.0 \pm 0.7) \, \si{\milli\meter} $ FWHM both in horizontal and vertical directions, while the IR spot was uniform in power and slightly astigmatic, $ (12.9 \pm 0.5) \times (17.2 \pm 0.5) \, \si{\milli\meter} $ FWHM in the horizontal and vertical directions, due to a geometrical cut of an optical element. \new{Two spurious light backgrounds at \SI{532}{\nano\meter} and \SI{894}{\nano\meter} were found superimposed to the \SI{1312}{\nano\meter} beam with the same spot size, caused by the non-ideality of the dichroic mirrors used to separate the beams, with energies $\approx \SI{80}{\micro\joule}$ and $\approx \SI{100}{\micro\joule}$ respectively. Photo-ionisation of \tS{} or \ttP{} due to these backgrounds was negligible due to their low intensities and the small photo-ionisation cross-sections ($ 10^{-16} \div 10^{-17} \, \si{\square\centi\meter} $).} As the presence of an electric field would have shorten considerably the optical lifetime of \tS{} Ps \cite{aegis_meta2:19}, a fast HV switch with a rise time $ < \SI{15}{\nano\second} $ was used to disable the guiding electric field \new{of the focusing electrode (see \fig{setpoint_comparison}.a)} $ \sim \SI{5}{\nano\second} $ after \pos{} implantation \cite{aegis_meta2:19}, such that the field was negligible when the excitation lasers were shot ($ \sim \SI{20}{\nano\second}$ after \pos{} implantation).}
\par\
\new{The time distribution  of the annihilation $ \gamma $ rays due to the implanted \pos{} and decaying/annihilating Ps was measured} with a $ 20 \times 25 \time 25 \, \si{\milli\meter} $ lead tungstate ($\mathrm{PbWO_4}$) scintillator \cite{cassidy_nimb:07} coupled to a Hamamatsu R11265-100 photomultiplier tube (PMT), placed \SI{40}{\milli\meter} above the Ps converter. The signal from the PMT was 50\%-50\% split and digitized using two channels of a HD4096 Teledyne LeCroy 2.5Gs oscilloscope set at high (\SI{100}{\milli\volt\per div}) and low (\SI{1}{\volt\per div}) gains to further extend the linear dynamic range of the digitizer. The data of the two channels were joined to form the so-called single-shot positron annihilation lifetime spectroscopy (SSPALS) spectra \cite{cassidy_sspals:06}, whose average is proportional to the amount of Ps/\pos{} annihilating per unit time. \new{The total lifetime of \tS{} Ps in absence of an electric field (\SI{1142}{\nano\second}) is longer than that of \otS{} (\SI{142}{\nano\second}) and of all other populated sublevels in the $n = 1-3$ manifolds. Thus, an increase in the delayed annihilations \new{(either due to in-flight or to pick-off annihilations as the Ps atoms hit the chamber walls)} for $ t \gg \SI{142}{\nano\second} $ ($ t $ being the time elapsed from \pos{} implantation) can be directly related to the amount of \tS{} Ps (as in \cite{aegis_meta2:19}).} Relative time-dependent variations between two SSPALS spectra families with different laser configurations were quantified using the parameter $ S(t) = (A_{1}(t) - A_{2}(t))/A_{1}(t) $ where $ A_{1}(t) $, $ A_{2}(t) $ are the (averaged) integrated areas below single SSPALS shots of the two families in a selected time window centered in $ t $. In this definition $ A_{1}(t) $ has the role of the reference area. An alternating measurement scheme between the two families was used to minimize the effect of time drifts of the experimental conditions. Contributions of eventual residual long-time drifts in the $ S $ calculation were further reduced by normalizing the shots in each family to a second-order polynomial fit of their value versus time (\textit{detrending technique}, see \cite{aegis_meta:18} App.). 

\new{First, a set of reference spontaneous \tS{} Ps production measurements was acquired on the detuned OPG set-point ($T_2$) to measure the $ S(t) = (A^\mathrm{off}(t) - A^\mathrm{UV}(t))/A^\mathrm{off}(t) $ parameter in the absence of stimulated emission in the conditions selected for the following experiments. The measured $S(t)$ curve was compared to one acquired at the on-resonance set-point ($T_1$) (see \fig{setpoint_comparison}), which was well characterized previously \cite{aegis_meta2:19}. $A^\mathrm{UV}$ and $A^\mathrm{off}$ correspond to the integrated averaged areas with and without the UV laser.}
Time windows of \SI{300}{\nano\second} width with steps of \SI{50}{\nano\second} were used to calculate $ S(t) $. All measurements were obtained alternating shots with the UV laser to the same number without it ($ \sim 200 $ shots for $ T_1 $, $ \sim 300 $ shots for $ T_2 $). The peak of \tS{} annihilations on the chamber walls is more evident in the on-resonance ($T_1$) measurement (between \SI{500}{\nano\second} and \SI{750}{\nano\second}) than in the detuned ($T_2$) measurement, where it is lower and smeared to longer times, in agreement with a reduced \tS{} production efficiency and longer atoms' traveled distances, because of the Doppler selection operated by the detuned UV beam that selects atoms traveling longer trajectories. 

The experimental $ S(t) $ curves were fitted using a previously-developed Monte Carlo (MC) model \cite{aegis_meta:18, aegis_meta2:19}. This MC calculates the atoms' flight trajectories (starting from our Ps source velocity distribution \cite{aegis_neq3:16}) and their excitation dynamics by simultaneously integrating the center-of-mass equations of motion and optical rate equations for the internal level dynamics. The annihilation position/time distribution for \otS{} and \tS{} Ps in flight and by collisions with the chamber walls \cite{aegis_meta:18} are calculated.  The only fitting parameter was the \otS{}--\ttP{} excitation efficiency $ \eta_3 $, while the \tS{} branching $ \eta_m $ ($ = 0.097$) and quenching $ \eta_q $ ($ = 0.17 $) efficiencies were set according to past measurements (with the same laser delay of \SI{20}{\nano\second}, see  \cite{aegis_meta2:19}). Different Doppler selections in the two set-points were now constrained to the measured UV detunings (\SI{0}{\giga\hertz} for $T_1$ and \SI{410}{\giga\hertz} for $T_2$). 

The fit (\fig{setpoint_comparison}, solid lines) yielded $ \eta_3 = (12.4 \pm 0.7) \% $ for $T_1$ and $ \eta_3 = (5.2 \pm 0.4) \, \% $ for $T_2$, \new{corresponding to a $ \approx \mkern-4mu 58 \, \% $ reduction in the excitation efficiency as a consequence of the UV laser detuning, in agreement with the expected $ \sim \mkern-4mu 50 \% $ \cite{aegis_neq3:16}}. The annihilations' position distributions  resulting from the fit (\fig{setpoint_comparison}.a) emphasize the difference in the atoms' flight trajectories in the chamber geometry in the two different detuning conditions.
\par
\begin{figure}[htp]
	\centering
	\includegraphics[width=0.9\textwidth]{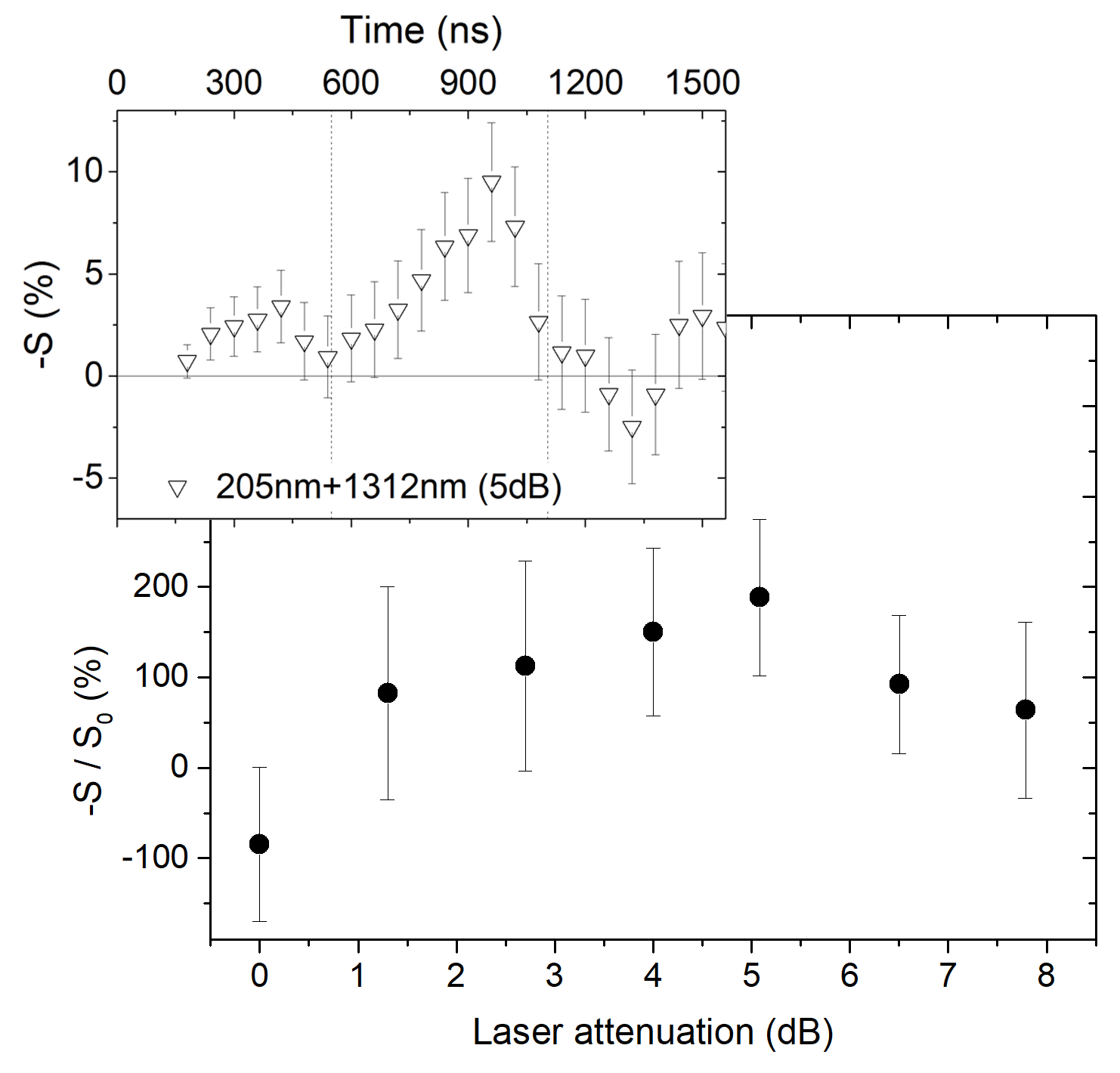}
	\caption{The ratio $ -\bar{S}/\bar{S}_0 $ between the total $ S $ parameters, as defined in the text, versus the attenuation in \si{\decibel} of the IR laser pulse starting from the maximum energy \SI{405}{\micro\joule} (black points): Top inset: the $ -S(t) $ parameter in the case of the \SI{5}{\decibel} attenuation point.}
	\label{fig:energy_scan}
\end{figure}

Subsequently, the \SI{1312.2}{\nano\meter} IR pulse was introduced to stimulate the \ttP{}--\tS{} transition. A first measurement campaign to optimize the IR laser \tS{} production efficiency was carried out by progressively attenuating its energy from the nominal \SI{405}{\micro\joule} with a set of graded neutral density filters. The idea behind this optimization search is that the desired stimulated emission from the \ttP{} level competes dynamically with the repumping of this level by absorption from the \tS{}. This mainly depends on the pulse energy (mantaining the other pulse parameters, in particular the FWHM temporal profile). One expects that on our nanosecond time scale, while a low energy can induce only a low population gain on the \tS{} level, a too high energy subtracts population from that previously efficiently excited by the rising part of the pulse. The optimization measurements were acquired alternating ($ \sim 200 $) shots with both UV and IR lasers to shots with UV laser only. Their associated $ S(t) = (A^\mathrm{UV}(t) - A^\mathrm{UV+IR}(t))/A^\mathrm{UV}(t) $ reflected the relative signal changes only caused by the IR, i.e. isolating the effects of this laser pulse (resonant with the \ttP{}--\tS{} transition) on the \tS{} Ps population. The resulting $ S(t) $ curves (an example of which - with \SI{5}{\decibel} attenuation - is shown in \fig{energy_scan}, inset) exhibit an excess of annihilations in the region around \SI{550}{\nano\second} and \SI{1050}{\nano\second}, where the impact of \tS{} atoms onto the chamber walls is also observed in the reference measurement ($T_2$ in \fig{setpoint_comparison}).
A convenient way to show the enhancement in the \tS{} production is to consider a total $ \bar{S} $ parameter calculated in the wide time window 550-1050 \si{\nano\second}, normalized on the total parameter $ \bar{S} $ of the reference measurement (calculated in the same time window). The results are reported in \fig{energy_scan} as a function of the \SI{1312.2}{\nano\meter} laser pulse attenuation. The effect of the \SI{1312.2}{\nano\meter} laser is compatible with zero when the beam is sent with full power. Then the effect progressively increases as the laser energy is decreased to reach a maximum at \SI{5}{\decibel} attenuation ($ E^{peak} \approx \SI{126}{\micro\joule} $) to slowly decrease again at even higher attenuations.

\new{As expected, with the maximum pulse energy a depopulation effect is observed. As the laser energy lowers, a positive gain in the \tS{} population progressively increases, reaching a maximum around the $ \SI{5}{\decibel} $ attenuation (where $ E_{peak} \sim  \SI{126}{\micro\joule} $), to slowly decrease again at even higher attenuations. These measurements indicate that, selecting the \SI{1312.2}{\nano\meter} laser pulse energy at $ E_{peak} $, an increase of $ (190 \pm 90) \% $ (corresponding to a $ \times (2.9 \pm 0.9) $ gain efficiency) in the amount of produced \tS{} with respect to the reference was obtained.}
\par
\begin{figure}[htp]
	\centering
	\includegraphics[width=\textwidth]{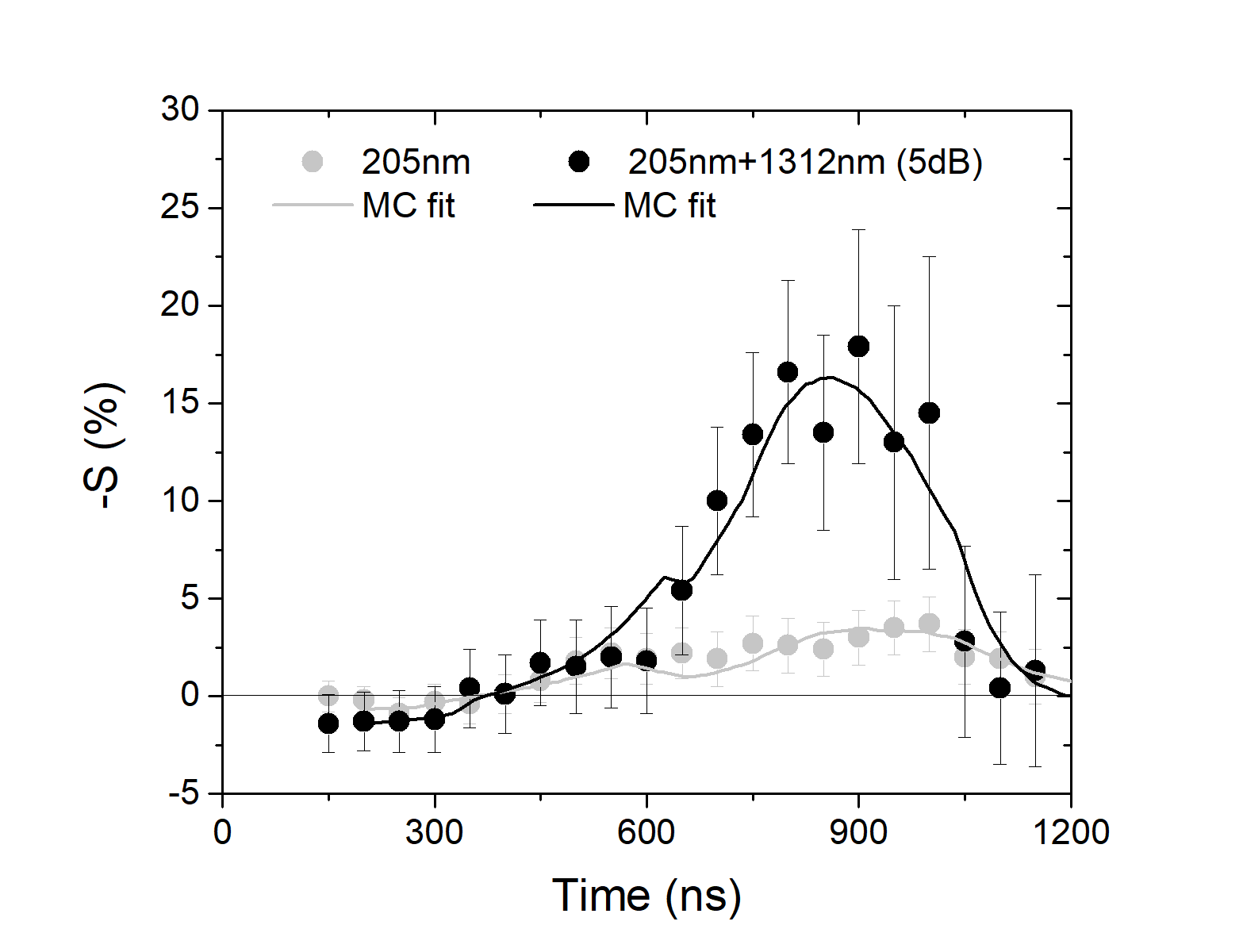}
	\caption{The experimental curve of the $ -S $ parameter ($ = (A^\mathrm{UV+IR}(t) - A^\mathrm{off}(t))/A^\mathrm{off}(t) $) measured in the empirically-determined optimal \SI{1312.2}{\nano\meter} laser intensity, compared to the reference $ -S $ obtained without the IR laser.}
	\label{fig:final_enhancement}
\end{figure}

A final measurement campaign was conducted to directly evaluate the \tS{} branching efficiency achieved by setting the \SI{1312.2}{\nano\meter} laser to the optimal energy. The set of measurements was acquired alternating ($ \sim \mkern-4mu 200 $) shots with both UV and IR lasers to shots with both lasers off.
The resulting $ S(t) = (A^\mathrm{off}(t) - A^\mathrm{UV+IR}(t))/A^\mathrm{off}(t) $ curve, compared to the reference \tS{} curve obtained in the same temperature set-point $ T_2 $ and in the absence of the IR enhancement laser (previously plotted in \fig{setpoint_comparison}), is shown in \fig{final_enhancement}. The \tS{} branching efficiency was evaluated by fitting $ -S(t) $ with the Monte Carlo model. The fit was now performed letting $ \eta_m $ vary, while keeping all other parameters fixed to those determined in the reference measurement. The found value for the branching efficiency was $ \eta^\mathrm{UV}_m = 0.297 \pm 0.019 $. This value, compared to the branching efficiency previously estimated from the measurement with the UV laser alone with the same laser delay ($ \eta^\mathrm{UV}_m = (0.097 \pm 0.027) $ \cite{aegis_meta2:19}), leads to a \tS{} Ps enhancement of $ \times (3.1 \pm 1.0) $, in perfect agreement with the IR pulse optimization measurement. 

In conclusion, we experimentally demonstrated the possibility to \new{efficiently stimulate the \ttP{}--\tS{} transition of Ps by employing} a pulsed, broadband \SI{1312.2}{\nano\meter} laser.
The highest enhancement efficiency was found by tuning the intensity of the IR laser pulse inducing the \ttP{}--\tS{} transition. In these optimal conditions of our experiment, due to the pulsed excitation dynamics, the relative enhancement of the \tS{} Ps atoms production from the excited \tS{} population, with respect to the reference of the spontaneous emission decay, was measured to be $ \times (3.1 \pm 1.0) $. This corresponds to a branching efficiency of \tS{} production from the \ttP{} level that can be up to $ \sim 30 \% $. Anyway, the overall \tS{} Ps excitation efficiency was limited by the present technical restriction in producing both pulsed laser beams with the correct wavelengths and energies using a single optical parametric generation stage. 
The variation in the temperature set-point of the laser generation crystal, necessary to output the correct \SI{1312.2}{\nano\meter} wavelength with sufficient energy, induced an UV detuning of about \SI{410}{\giga\hertz} and a reduction in the amount of excited \ttP{} of $ \approx \mkern-4mu 58 \, \% $.
\par
In a future realization of this experiment, the present technical limitations could be overcome by separating the two UV and IR laser lines, i.e. having independent non-linear optical generation and amplification stages. An advantage of having the two laser wavelengths independent from each other would be to retain the full tunability characteristics of our \tS{} source \cite{aegis_meta2:19}, \new{while conveying $ \sim 30 \% $ of what could be excited to \ttP{} to \tS{}. Furthermore, if one accepts to sacrifice the mentioned velocity selection, a further increase up to a factor of five of the overall excitation efficiency could be obtained by enlarging the UV laser bandwidth to cover efficiently the Ps Doppler profile. Finally, the first laser spectroscopy of the \ttP{}--\tS{} transition would become feasible. A laser system with independent UV and IR laser lines is currently under development to take full advantage of this \otS{}--\ttP{}--\tS{} stimulated excitation scheme in view of future measurements on a beam \tS{} Ps.}

\vspace{0.5cm}

The authors are grateful to Dr. S. Cialdi for the original development of the \otS{}-\ttP{} laser. This work was supported by Istituto Nazionale di Fisica Nucleare; the CERN Fellowship programme and the CERN Doctoral student programme; the Swiss National Science Foundation Ambizione Grant (No. 154833); a Deutsche Forschungsgemeinschaft research grant; an excellence initiative of Heidelberg University; Marie Sklodowska-Curie Innovative Training Network Fellowship of the European Commission's Horizon 2020 Programme (No. 721559 AVA); European Research Council under the European Unions Seventh Framework Program FP7/2007-2013 (Grants Nos. 291242 and 277762); European Union's Horizon 2020 research and innovation programme under the Marie Sklodowska-Curie grant agreement ANGRAM No. 748826; Austrian Ministry for Science, Research, and Economy; Research Council of Norway; Bergen Research Foundation; John Templeton Foundation; Ministry of Education and Science of the Russian Federation and Russian Academy of Sciences and the European Social Fund within the framework of realizing the project, in support of intersectoral mobility and the European Social Fund within the framework of realizing Research infrastructure for experiments at CERN, LM2015058.

\bibliographystyle{apsrev}
\bibliography{aegis_biblio}

\end{document}